\pdfoutput=1   % needed for arXiv, and should be within the first 5 lines of the file!!!

\documentclass[useAMS]{mn2e}
\usepackage{amsmath}
\usepackage{textcomp}
\usepackage{amssymb}
\usepackage[pdftex]{graphicx}
\usepackage{amssymb}

\usepackage[margin=.8in, paperwidth=8.5in, paperheight=11in]{geometry}
\usepackage{multicol}
\usepackage{multirow}
\usepackage{epsfig}
\usepackage{graphicx}
\usepackage{dblfloatfix}
\def\lesssim{\mathrel{\hbox{\rlap{\hbox{\lower4pt\hbox{$\sim$}}}\hbox{$<$}}}}
\def\gtrsim{\mathrel{\hbox{\rlap{\hbox{\lower4pt\hbox{$\sim$}}}\hbox{$>$}}}} 
\usepackage{accents}
\newcommand*{\dt}[1]{%
  \accentset{\mbox{\Large\bfseries .}}{#1}}
\title[Stellar Wind Erosion of Protoplanetary Discs]{Stellar Wind Erosion of
Protoplanetary Discs}
\author[N. R. Schnepf, R. V. E. Lovelace,  M.M. Romanova, \&
V.S. Airapetian]{N. R. Schnepf$^{1}$,  R. V. E. 
Lovelace$^{2}$,  M.M. Romanova$^2$,
and V.S. Airapetian$^3$\\
$^{1}$Earth, Atmospheric and Planetary Sciences, MIT, Cambridge, 02139 MA, USA: email: nschnepf@mit.edu,\\
$^{2}$Department of Astronomy, Cornell University, Ithaca, 14853 NY, USA: email:lovelace@astro.cornell.edu; romanova@astro.cornell.edu,\\
$^3$NASA/Goddard Space Flight Center, Greenbelt, MD 20771,
USA: email: vladimir.airapetian-1@nasa.gov}

\begin{document}

\maketitle

\begin{abstract}

    An analytic model is developed for the erosion of protoplanetary gas discs by high velocity magnetized stellar winds.   The winds are centrifugally driven from the surface of  rapidly rotating, strongly magnetized young stars.       The presence of the magnetic field in the wind leads to Reynolds numbers sufficiently large to cause a strongly turbulent wind/disk boundary layer which entrains and carries away the disc gas.   The model uses the conservation of mass and momentum in the turbulent boundary layer.    The time-scale for significant erosion depends on the disc accretion speed, disc accretion rate, the wind mass loss rate, and the wind velocity.  The time-scale is estimated to be $\sim 2\times 10^6$ yr.    
       The analytic model assumes a steady stellar wind with mass loss rate
 $\dt{M}_w \sim 10^{-10}M_\odot$ yr$^{-1}$ and velocity $v_w \sim
 10^3 $~km s$^{-1}$.       A significant contribution to the disc erosion 
 can come from frequent  powerful coronal mass ejections (CMEs) where the average mass loss rate in CMEs, $\dt{M}_{\rm CME}$, and velocities, $v_{\rm CME}$, have values comparable to those for the steady wind.

\end{abstract}


\section{Introduction} 

  A wide body of observations establish that the lifetime of gaseous protoplanetary discs is relatively short.  While most
  protostars younger than $10^6$ yr have gaseous discs, 
  only $50\%$  of protostars $3\times 10^6$ yr in age have  gaseous discs and very few stars $6\times 10^6$ yr   or older have gaseous discs (Armitage 2010).    From spectroscopic observations of the hot continuum radiation produced when infalling gas impacts the stellar surface, it is known that the accretion rate of gas onto a star decays on a similar time-scale (Hartmann et al. 1998). Observations also suggest that the dispersal of gas occurs on a wide range of disc radii during a short time-scale (Skrutskie et al. 1990; Wolk \& Walter 1996;  Andrews \& Williams 2005). 

The short lifetime of a gaseous disc plays a critical role in the formation of planetary systems. It directly affects the time available for  planetesimals to agglomerate  additional material, as well as the migration of planets within the disc (Armitage 2010;  Zsom et al. 2010). The mechanisms by which the gas is lost may play an important role in the formation of planetary systems.

   A class of protoplanetary discs termed {\it transition discs} have
been identified by a dip in the mid-infrared spectra which can be modeled
by a reduced surface density of dust in the inner regions ($\lesssim 35$ au) of the discs (Espaillat et al. 2014; Espaillat et al.. 2007).
  Sub-millimeter imaging  directly shows a deficit in surface density
of dust in these systems (e.g., Pi\'etu et al. 2007).  The observations
suggest an inside-out dispersal of the dust possibly caused by
photo-evaporation (Clarke et al. 2001; Owen et al. 2012), by
the presence of giant planets, or other processes (Birnstiel et al. 2013).  
  Photo-evaporative destruction of discs is thought to occur when
the photo-evaporative mass-loss rate $\dt{M}_{\rm pe}$ exceeds the accretion rate  $\dt{M}_d$ onto the star
 (e.g.,  Richling \& Yorke 1997; Hollenbach et al. 2000).  
   Additionally, the radial distribution of dust is influenced by grain growth, migration, destruction in collisions, and trapping in vortices excited at the  outer gap edge caused by giant planets (Reg\'aly et al. 2012).
     Transition discs have lower accretion rates than younger
standard discs, but not sufficiently low to explain the  large sizes of
the regions of low dust density by photo-evaporation (Birnstiel 2013).
    A promising possibility is that  the depletion of dust in the inner regions of 
 transitional disc is due to the presence of multiple planets located
 at radii $\sim 0.1 - 10$ au (Espaillat et al. 2014).

%%%%%%%%%% FIGURE 1 %%%%%%%%%%%%%%%%%%%
	\begin{figure}
	\includegraphics[width=0.49\textwidth]{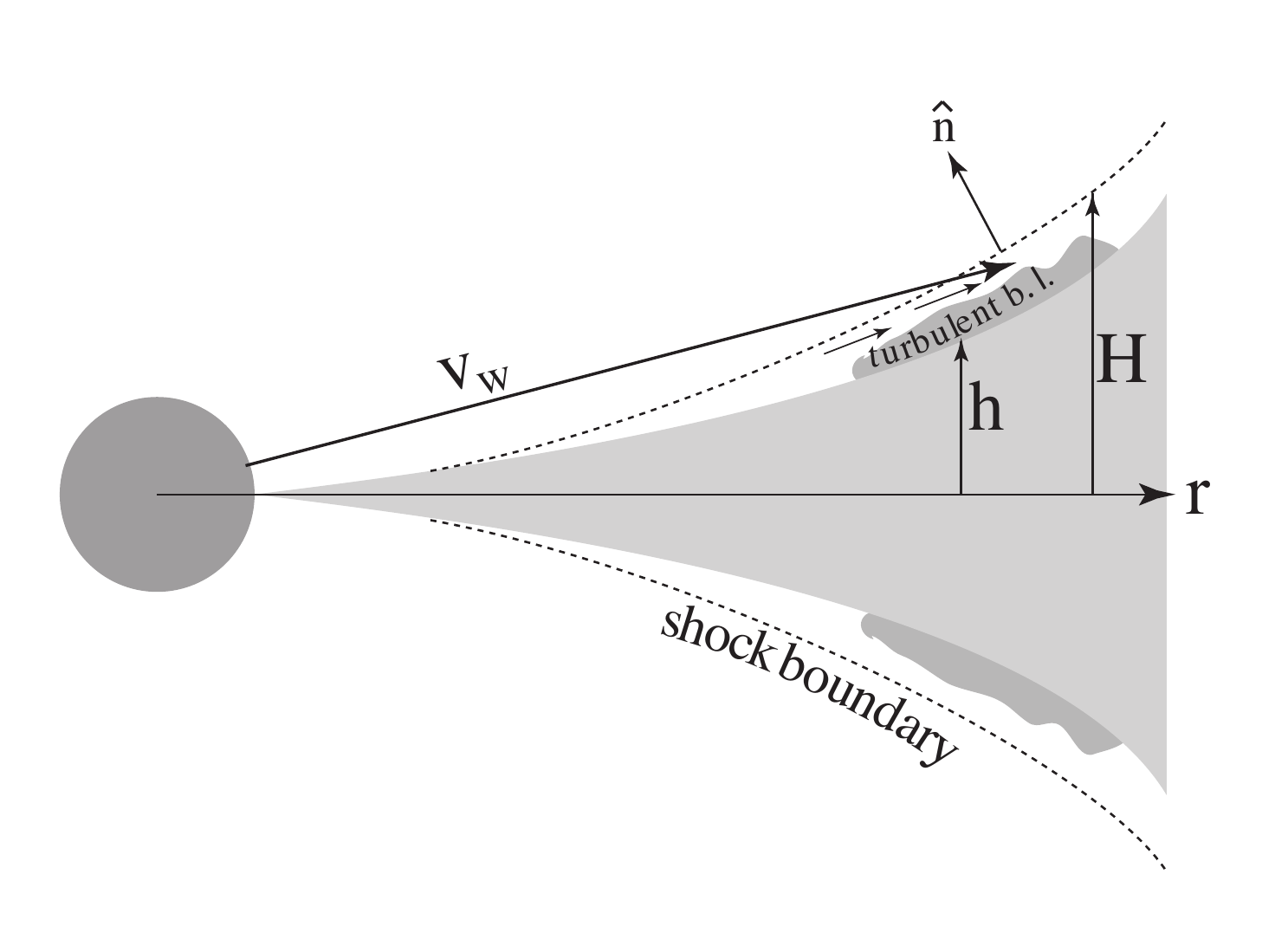}
	\caption{A cross-section of the protoplanetary disc with the disc-wind turbulent boundary layer ($h<z<H$) and outer oblique shock shown. $\mathbf{\hat{n}}$ is the normal to the shock wave, $\mathbf{v}_w$ is the wind velocity, $h$ is the disc half-thickness and $H$ is the vertical height of the shock.  The figure has been adapted from a figure in
Lovelace et al. 2008. }
	\label{DiscGeometry}
	\end{figure}
%%%%%%%%%%%%%%%%%%%%%%%%%%%%%%%%%%%

       Multiple processes  may be responsible for the loss
of the gas and entrained dust from protoplanetary discs.   
      This work analyzes  the erosion of the gas and entrained dust
particles from discs caused by high-velocity magnetized stellar winds.  
   The presence of the magnetic field leads to Reynolds numbers sufficiently large to cause a strongly turbulent wind and disk boundary layer (see Figure \ref{DiscGeometry}) as suggested by Lovelace,
   Romanova, \& Barnard (2008).    
      Strong magnetized winds from young stars with discs are likely 
because the stars are known to rotate rapidly and  to be strongly magnetized. The magnetic winds depend on both the star's rotation rate {\it and} its magnetic field owing to complex and likely random dynamo processes. Thus a one-to-one correlation with the rotation rate is not expected.
      Wind erosion of protostellar discs has been discussed in a number of previous works as reviewed by Hollenbach et al. (2000).
   However, the important role played by a rapidly rotating star's magnetic field in launching a high velocity wind   has apparently  not been considered.
   Furthermore, the role of the wind's magnetic field in producing a strongly turbulent wind/disc boundary layer  has not been discussed.    
    In addition to a steady
 high-velocity stellar wind, we consider the disc erosion due to frequent powerful coronal mass ejections.

     Section 2 of the paper discusses magnetized stellar winds.
      Section 3  develops an analytic model for the evolution
of the mass surface density of the disc $\Sigma_d(r,t)$.      Section 4
discusses the contribution to  disc erosion resulting from  frequent 
powerful coronal mass ejections.
 Section 5 gives the conclusions of this work. 

\section{Magnetized Stellar Winds}

    The winds from rotating
magnetized stars may be thermally
driven as in the Solar wind.
   The magnetic field has an important
role in the outward transport of angular momentum by the wind 
(Weber \& Davis 1967; Matt \& Pudritz 2008a,b).
     For thermally driven winds from slowly
rotating stars,  the radial flow
velocity at  large distances is the Parker velocity 
$
v_P = [2 c_{s0}^2/  (\gamma-1) -2 G M_*/ R_*]^{1/2}~,
$
where $c_{s0}$ is the sound speed at the star's radius $R_*$,
$M_*$ is its mass, and $\gamma$ is the specific heat
ratio.  [Note that both spherical $(R,\theta,\phi)$  and
cylindrical $(r, \phi,z)$ inertial coordinates are used in this
work.]

    In contrast, for conditions where 1) the thermal
speed of the gas close to the star is small (compared
with the velocity), 2) the star's magnetic field
is strong, and 3) the star rotates rapidly,
there are magnetically driven winds (also termed fast magnetic
rotator, FMR,  winds) (Michel 1969; Belcher \& MacGregor 1976). 
    These winds are driven by the centrifugal force
of the star's rapidly rotating magnetic
field rather than the thermal energy in the star's corona.

    For a rotating magnetized star, the radial wind velocity
at large distances from the star is  
$v_w =1.5[ \Omega_*^2 (B_{R*}R_*^2)^2 
/ \dt{M}_w]^{1/3}$
 (Michel (1969), where  $P_*=2\pi/\Omega_*$ is the
rotation period of the star and
$B_{R*} =  B_R(R_*)$.   
    The  characteristic acceleration
distance of the winds is the Alfv\'en radius of the wind,
$R_{Aw} =(2/3)^{1/2}(v_w/\Omega_*)$ (Belcher \& MacGregor 1976).
        For a star with rotation period $P_*=10$ d,  a radius
twice the radius of the Sun, $R_*=1.4\times 10^{11}$ cm
(Armitage \& Clarke 1996),  a surface magnetic
field $B_{R*}= 0.5$ kG, and $\dt{M}_w =10^{-10}M_\odot$/yr,   one finds $v_w \approx 1400$ km/s, and $R_{Aw} \approx 1$ au.   
    For the magnetic acceleration of the solar wind to be important,
the product of the rotation rate $\Omega_*$ times the magnetic
flux per sterradian $B_{R*} R_*^2$ would need to  be $\sim 20$
times larger than the present values (Belcher \& MacGregor 1976).
    The magnetic winds cause the rapid spin-down of the
stars on a time-scale $T_{sdw} = I[(2/3) R_{Aw}^2 \dt{M}_w]^{-1}$,
where $I$ is the moment of inertia of the star
(Weber \& Davis 1967; Belcher \& MacGregor 1976).
   This time-scale may be as short as $\sim 10^6$ yr.  However,
the spin-down torque of the wind is strongly dominated by
the spin-up torque due to the disc
accretion to the star for accretion rates $\sim 10^{-8}M_\odot$/yr.
   In this regime the star's rotation tends to be locked to the rotation 
rate of the inner disc by the magnetic coupling between the 
star and the disc as proposed by K\"onigl (1991) and observed
in MHD simulations by Long et al. (2005).   In contrast, Matt and
Pudritz (2008a,b) argue theoretically that the magnetic coupling
is ineffective and that an accretion powered stellar wind acts to
counteract the tendency of the accretion to spin-up the star.
   A review of this topic is given by Bouvier et al. (2014).

   The magnetic fields of classical T
Tauri stars (CTTSs) are typically 
in the kG range (Johns-Krull
\& Valenti 2000; Johns-Krull 2007)).  
Furthermore,  these stars typically rotate 
rapidly with periods $\sim 2-15$ d (Bouvier et al. 2014; Bouvier et al. 1993). 
   Thus magnetically driven winds may be important for T Tauri
stars.

 Magnetically driven winds from  the disc/magnetospheric
 boundary have been found in magnetohydrodynamic (MHD)
simulations of rapidly-rotating, disk-accreting stars particularly in the ``propeller" regime (Romanova et al. 2005; Ustyugova et al. 2006; Lovelace et al. 1999; Romanova et al. 2009;  Lii et al. 2012; Lii
et al. 2014).   The propeller regime arises when the magnetospheric
radius $r_{\rm m}$ becomes  larger than the corotation radius
$r_{\rm cr} =(GM_*/\Omega_*^2)^{1/3}$.  Because $r_{\rm m}$
depends inversely on the disc accretion rate $\dt{M}_a$ (to a 
fractional power), the propeller regime is unavoidable  as $\dt{M}_a$
decreases in the late stages of disc evolution.
  The winds are found to flow in opposite directions
along the rotation axis transporting energy and angular
momentum away from the star.
    For the case of an aligned a dipole field, the field
acts to block an equatorial outflow.  However,
for  multipolar fields, winds are expected in all
directions from the star's surface.

In deriving the evolution of disc density due to this magnetized wind, it is useful to consider the physical conditions in the wind at a distance from the sun of $R=1$ au with fiducial conditions of wind density $n_w=10^5$ cm$^{-3}$, wind speed $v_w=10^3$ km/s, a predominantly toroidal time averaged magnetic field $B_w=0.1$ G, and ion (proton) and electron temperatures of $T_i=T_e=10^5$ K. From this distance and beyond, 1) the wind velocity is predominantly radial, 2) it is super fast magnetosonic, and 3) it is much larger than the Keplerian velocity of the disc matter. 

Under these conditions, the ion and electron gyro-frequencies are $\omega_{ci}\approx 10^3$/s and $\omega_{ce}\approx 2\times10^6$/s. The corresponding ion and electron gyro-radii are $r_{gi}\approx3\times10^3$ cm and $r_{ge}\approx 70$ cm. From this, the ion and electron collision times ($\propto T^{3/2}/n$) are $\tau_i\approx 470$ s and $\tau_i\approx 11$ s. The ion and electron mean-free paths are $\ell_i=\ell_e=1.4\times10^9$ cm (Braginskii 1965). Thus, $\omega_{ci}\tau_i\approx4.6\times10^5$ and $\omega_{ce}\tau_e\approx2\times10^7$.

In the absence of a magnetic field, the kinematic viscosities of ions and electrons are $\nu_{0i}\approx v_{thi}^2\tau_i\approx4\times10^{15}$ cm$^2$/s and $\nu_{0e}\approx v_{the}^2\tau_e\approx 2\times10^{17}$ cm$^2$/s, where $v_{{\rm th},i,e}$ is the ion or electron thermal speed. Thus, without a magnetic field the Reynolds number, ${\rm Re}=rv_w/\nu_{0e}\approx 9000$, is such that a boundary layer flow is laminar. 

When the magnetic field is included, there are five different viscosity coefficients. Fortunately, for the considered problem the important viscosity coefficient is that for momentum transport across the magnetic field. For example, the momentum flux-density component is $T_{R\theta}=-\rho\nu_{\perp}R^{-1}\partial v_R/\partial\theta$. For ions, the viscosity term is $\nu_{\perp i}\approx \nu_{0i}/(\omega_{ci}\tau_i)^2\approx 2\times10^4$ cm$^2$/s, and for electrons it is $\nu_{\perp e}\approx \nu_{0e}/(\omega_{ce}\tau_e)^2\approx500$ cm$^2$/s. 
    Using the viscosity $\nu_{\perp i}$, the effective Reynolds number for the wind is ${\rm Re}_w=Rv_w/\nu_{\perp i} \sim 10^{18}$. Evidently, unlike the non-magnetic case with laminar boundary layer flow, with a magnetic field the boundary layer flow is strongly turbulent. The large reduction of the viscosity results from the particle step size between collisions being a gyro-radius rather than a mean-free path. Thus, the estimated Reynolds number also holds for a turbulent magnetic field.

The relevant heat conductivity coefficient is that for the heat flux across the magnetic field, $q_\theta=-\kappa_\perp R^{-1}\partial(k_B T)/\partial \theta$, where $\kappa_\perp\approx\kappa_{\perp i}\approx2nv_{thi}^2\tau_i/(\omega_{ci}\tau_i)^2\approx3.8\times10^9$ (cm s)$^{-1}$ and where $k_B$ is Boltzman's constant. For this heat conductivity, the heat flow from the wind into the disc is negligible compared with the energy flux per unit area from the disc [ $3GM_*\dt{M}_d/(8\pi R^3)$ ] for accretion rates $\dt{M}_d \sim10^{-8}$M$_\odot$/yr, which is
in turn small compared to the irradiation of the disc by the star.

A weak oblique shock arises where the wind encounters the much denser disc. This occurs at a height $H(r)\ll r$ above the equatorial plane for a disc with half-thickness $h(r)<H(r)$ (Fig. \ref{DiscGeometry}). The angle between the incident flow and the shock is $\beta\approx r \mathrm{d}(H/r)/\mathrm{d}r\ll1$, with $\beta$  expected to be $>0$. 

After passing through the shock, the flow is deflected by an angle $\delta=2\beta/(1+\gamma)=3\beta/4$ (where $\gamma=5/3$) away from the equatorial plane, and the flow speed is reduced by a small fractional amount. This region between $h$ and $H$ is the boundary layer. The influx of wind matter into this layer is $-\mathrm{d}\mathbf{S}\cdot(\rho_w \mathbf{v}_w) \approx \mathrm{d}S\rho_wv_wr[\mathrm{d}(H/r)/\mathrm{d}r]$, where $\mathrm{d}S = r\mathrm{d}r\mathrm{d}\phi$ is the area element of the shock on the top side of the disc. The Keplerian velocity ($v_K=\sqrt{GM/r}$) of the disc is small compared to the wind velocity for $r\ge1$ au, so it is neglected. The density in the boundary layer varies from $\rho(r,h)\gg\rho_w$ at the surface of the disc to $\rho_w$ at $z=H$.  The time-averaged radial flow velocity varies from $|v_r(r,h)|\ll v_w$ to $v_r=v_w$ at $z=H$; $v_r(r,h)$ is neglected. 

%%%%%%%%% FIGURE 2 %%%%%%%%%%%%%%%%%%%%
	\begin{figure}
	\includegraphics[width=0.49\textwidth]{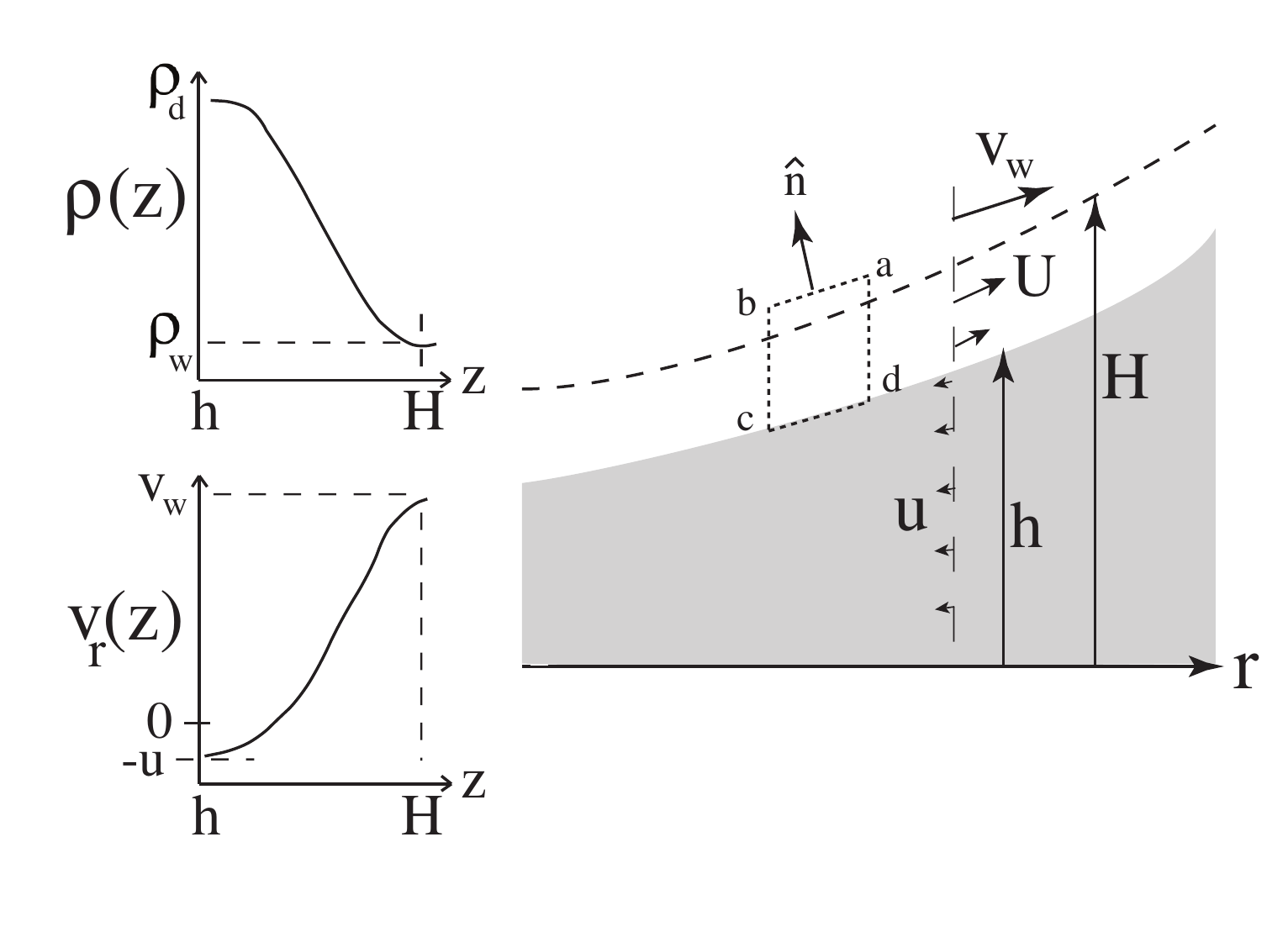}
	\caption{Sketches of the vertical profiles of density $\rho$ and radial velocity $\langle  v_r\rangle$ within the boundary layer is shown on the left. Here, $\rho_d$ is the density at the surface of the disc, $\rho_w$ is the density of the wind, and $v_w$ is the speed of the wind.  
	 The figure has been adapted from a figure in
Lovelace et al. 2008.}
	\label{BL_density}
	\end{figure}
%%%%%%%%%%%%%%%%%%%%%%%%%%%%%%%%%%%

      For an explanation of our notation it is useful to first consider
stationary conditions (later we evaluate the time evolution of the
disc).  In this case, the conservation of mass and radial momentum in the annular region $a-b-c-d$ (shown in Fig. \ref{BL_density}) of the boundary layer on the top side of the disc provides 
\begin{equation}
	\frac{\partial\left( rF_r^m \right)}{\partial r} = r \left[  r \frac{\mathrm{d}(H/r)}{\mathrm{d}r}\right]\rho_w v_w + \frac{1}{4\pi}\frac{d\dt{M}_d}{dr}~,
\end{equation}
\begin{equation}
	\frac{\partial\left( rF_r^p \right)}{\partial r} = r \left[  r \frac{\mathrm{d}(H/r)}{\mathrm{d}r}\right]\rho_w v_w^2~.
\end{equation}
 The terms on the left-hand side are due to the vertical sides of the region and the right-hand side is due to the sides $a-b$ and $c-d$. Here, $F_r^m$ is the mass flux and $F_r^p$ is the radial momentum flux. Both fluxes are given per unit circumference of the top side of the disc,
\begin{equation}
	F_r^m=\int_h^H dz \langle \rho v_r\rangle,~~~~F_r^p=\int_h^H dz 
	\langle \rho v_r^2\rangle~.
\end{equation}
  The average radial velocity
in the boundary layer is 
\begin{equation}
U(r) =
{F_r^p(r)  \over F_r^m(r)} ~.
 \end{equation}
The angular brackets indicate averages are over the turbulent fluctuations. The mass-loss rate of the disc per unit radius due to entrainment is d$\dt{M}_d/$d$r$. The disc matter influx to the boundary layer brings in negligible radial momentum.

The average velocity $U$ depends on the vertical profiles of density and radial velocity-- which are both unknown.
   We expect the profiles, for example
$\langle v_r(z)\rangle$ to be
substantially different from those of
laboratory turbulent boundary layers
over solid surfaces where $\rho=$ const (e.g., Schlicting  1968;
Roy \& Blottner 2006).
The main reason for the difference is
that the  density at the
surface of the disc $\rho(r,h)$
is many orders of magnitude larger 
than the wind density $\rho_w$.
  For a laboratory
boundary layer, a mixing-length model of the
momentum transport gives $(z^\prime)^2
(d\langle v_r\rangle/dz^\prime)^2=$ const, where
 $z^\prime \equiv z-h$, and this gives
the well-known logarithmic velocity profile
(see e.g. Schlicting 1968).
  For this profile most of the change
of velocity is quite close to the wall ($z^\prime =0$).
  In contrast, for the disc boundary
layer  a mixing
length model  gives $\rho(z^\prime)(z^\prime)^2
(d\langle v_r\rangle/dz^\prime)^2=$ const.
Because of the density dependence, the change in
the velocity  occurs  relatively far from 
from the wall.
   An important consequence of this is that
the average velocity $U$ is much smaller
than $v_w$, because $U$ is the density weighted
average radial velocity.   Here, we assume
that $U$ is larger by a factor of $g >1$
than the local escape velocity 
$v_{\rm esc}=(2GM_*/r)^{1/2}$ so that
 the  matter flow in the boundary
layer escapes  the star.

\section{Non-Stationary Evolution}

     In the absence of wind erosion, 
conservation of the disc matter gives
\begin{equation}
	{\partial(2\pi r\Sigma_d )\over\partial{t}}
	-{\partial ( 2{\pi}r u\Sigma_d)\over\partial{r}}=0~,
\end{equation}	
where 
\begin{equation}
\Sigma_d(r,t) = \int_{-h}^h dz~\rho(r,z,t)~,~~~
u(r,t) =-{1\over \Sigma_d}\int_{-h}^h dz~\rho v_r~,
\end{equation}
with $u \geq 0$ being the accretion speed of the disc matter.

     For stationary conditions, $r u \Sigma_s=$ const.  We
assume an $\alpha-$disc model (Shakura \& Sunyaev 1973): 
$u= \alpha c_s^2/v_K$, where $\alpha=$ const $ \sim 10^{-3}-0.1$, $c_s$ is the midplane isothermal sound speed, and $v_K$ is the 
Keplerian velocity of the disc.   Commonly considered
models have $\Sigma_d \propto r^{-q}$ with $q=$ const $\sim 1$.
This implies that $T \propto r^{q-3/2}$ and  $u \propto r^{q-1}$.
  The disc is assumed to be in hydrostatic equilibrium in
  the vertical direction so that
  $h/r = c_s/v_K$ and thus $h/r \propto r^{q/2 -1/4}$.   For specificity, we adopt $q=1$ so that $u=$ const and  $h/r \propto r^{1/4}$.

     With the wind erosion included, we consider that $u$
and $h/r$ are the same as in a stationary disc in
the region where  $\Sigma_d(r.t) >0$.   The basis for this
is the assumption of the Shakura and Sunyaev (1973) $\alpha=$  const
model  where  $u$ and $h$ depend on the midplane disc 
temperature which in turn depends on the stellar irradiation of the
 disc.
Further, we
   assume that the surface density in the top and bottom boundary layers,
$\Sigma_{\rm bl}=2\int_h^Hdz~\rho$, is much smaller than $\Sigma_d$.
  Mass conservation then gives 
\begin{multline}
		\frac{\partial\left(2{\pi}r\Sigma_d \right)}{\partial{t}}
	-	\frac{\partial\left( 2{\pi}ru\Sigma_d\right)}{\partial{r}}\\
	=-4{\pi}\frac{\partial \left( r{F_r}^m\right) }{\partial{r}}
		+4{\pi}r^2\left[\frac{d}{dr}\left( \frac{H}{r}\right)\right]{\rho_w}v_w
\end{multline} 
where ${F_r}^m$ is given in Eqn. 3.

Conservation of momentum in the radial direction gives
	\begin{equation}
	\frac{\partial \left(2\pi r \Sigma_{\rm bl} U \right)}{\partial{t}} +\frac{\partial 
	\left( 4{\pi}r{F_r}^p\right)}{\partial{r}}
	=4{\pi}r^2\left[\frac{d}{dr}\left(\frac{H}{r} \right)\right]\rho_w{v_w}^2 ~,
\end{equation}
where $F_r^p$ is given in Eqn. 3 and  $U(r,t)$ in Eqn. 4.   The term 
$\partial (2\pi r \Sigma_{\rm bl}U)/\partial t$ can be neglected because
$\Sigma_{\rm bl} \ll \Sigma_d$.   
   Equation 8 can then be integrated from the inner radius
of the disc $r_{\rm in}$ where $F_r^m=0$  to a radius $r$ to give
\begin{equation}
rF_r^m ={\dt{M}_w v_w \over 4 \pi U}\left[{H\over r}
-\left({H \over r}\right)_i \right]~,
\end{equation}
where the $i-$subscript indicates evaluation at $r=r_{\rm in}$.  
We have again assumed for simplicity that the stellar
wind is spherically symmetric with both $\dt{M}_w=$ const and $v_w=$ const in space and time for $t>0$.
    Thus the right-hand side of Eqn. 7 is
 \begin{equation}
 {\cal R}_{10} =-\dt{M}_w{\partial \over \partial r}
 \left({ v_w \over U} \left[{H\over r }- \left({H\over r}\right)_i\right]-{H \over r}\right)~.
  \end{equation}

     Equation 7 can be integrated using the method of 
 characteristics.   In view of the fact that $u=$ const, we
 have 
 \begin{equation}
 {d (2\pi r \Sigma_d)\over dt } =
 \left({\partial \over \partial t} -u{\partial \over \partial r}\right)(2\pi \Sigma_d)~.
 \end{equation}
 The characteristics are $r=r_0 - ut$, where $r_0$ is the radius
 of the disc fluid element at $t=0$ with $r_{\rm in} \leq r_0 \leq
 r_{\rm out}$ and $r_{\rm out}$ the outer radius of the disc.
    At $t=0$ the surface density of the disc is $\Sigma_{d0} =
 \Sigma_{0i}(r_{\rm in}/r)$.  Hence
 \begin{equation}
 {d\hat{\Sigma}_d \over dt} ={{\cal R}_{10} \over
 2\pi r_{\rm in} \Sigma_{0i}}~,
 \end{equation}
 where $\hat{\Sigma}_d(r,t)  \equiv \Sigma_d(r,t)/\Sigma_{d0}(r)$.
   Integrating this equation gives
\begin{eqnarray}
\hat{\Sigma}_d(r,t)&=& 1+\int_0^t dt~ {{\cal R}_{10} 
\over 2\pi r_{\rm in}\Sigma_{0i}}~,\nonumber \\
&=&1+\int_r^{r+ut} dr^\prime ~{{\cal R}_{10}(r^\prime)\over
2\pi r_{\rm in}u \Sigma_{0i}}~, \nonumber \\
&=& 1-F(r+ut)+F(r)~,
\end{eqnarray}
where 
\begin{equation}
F(r) ={\dt{M}_w \over \dt{M}_{d0}}\left({v_w \over U(r)}
\left[{H\over r} -\left({H\over r}\right)_i\right] -{H\over r} \right)~,
\end{equation}
 with $\dt{M}_{d0} =2\pi r_{\rm in} \Sigma_{d0}=$ const being the
 disc accretion rate at the inner radius of the disc assuming the
 disc extents into $r_{\rm in}(t=0)$.     As discussed in Sec. 3, we consider 
 the mean boundary layer velocity $U$ to be a factor $g\geq 1$
 times larger than the local escape velocity; that is, 
 $U(r)=g (2 G M/r)^{1/2}$.   Additionally, we assume
 $H/r = (H/r)_i (r/r_i)^\beta$ with $\beta > 1/4$.

  %%%%%%%%% FIGURE 3 %%%%%%%%%%%%%%%%%%%%
	\begin{figure}
	\includegraphics[width=0.49\textwidth]{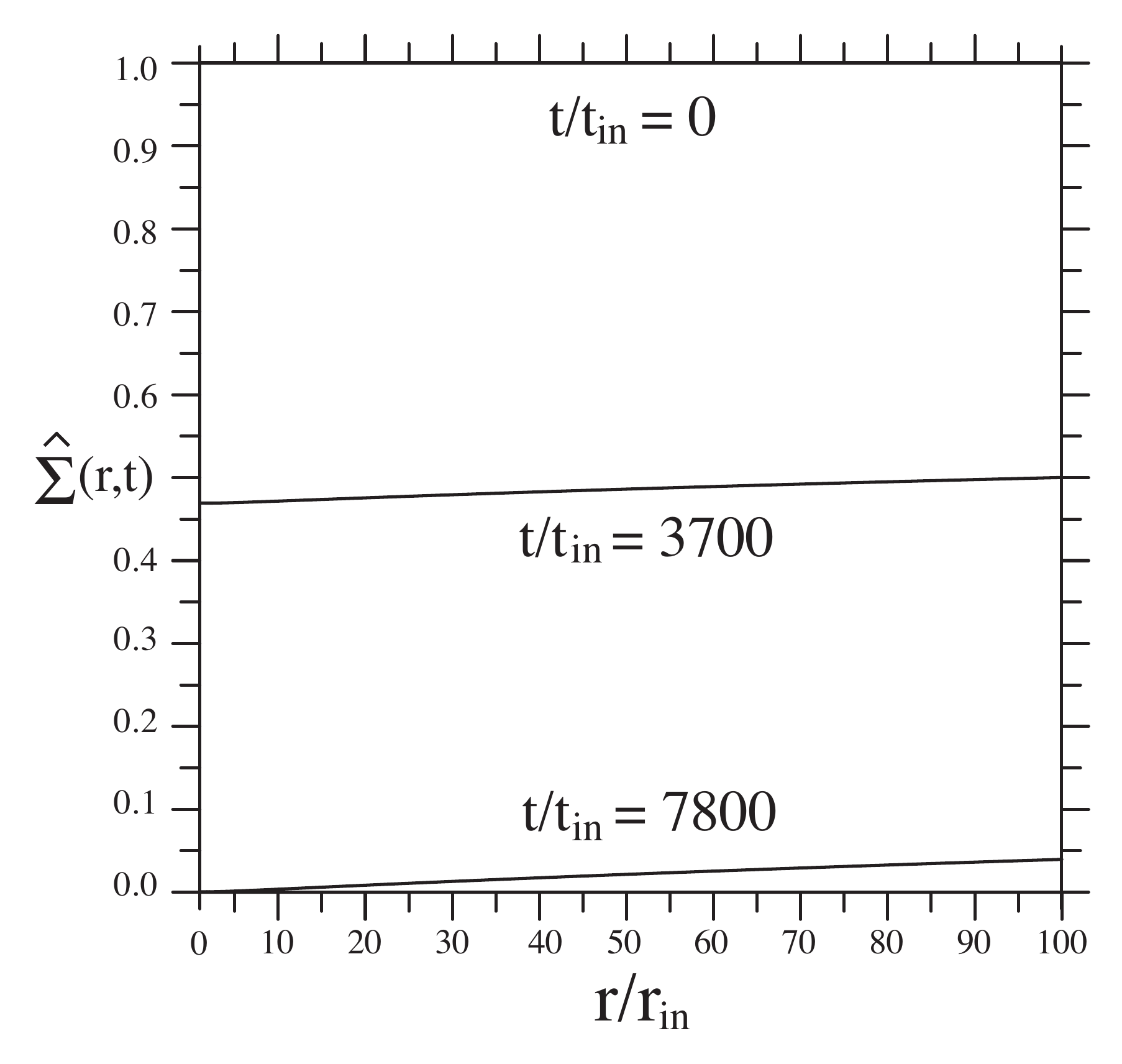}
	\caption{ Evolution of the disc surface density 
$\hat{\Sigma}_d(r,t)=\Sigma_d(r,t)/\Sigma_d(r,0)$ for an illustrative
case discussed in the text.   The radius is measured in
units of the initial  inner radius of the disc, $r_{\rm in}(t=0)$, which
is taken to be $10r_\odot=7\times 10^{11}$ cm.  Time is
measured in units of $t_{\rm in} = r_{\rm in}/u =257$ yr, where
the accretion speed is $u=86.3$ cm/s corresponding to a viscosity
coefficient $\alpha = 10^{-2}$ and a disc half-thickness at $r_{\rm in}$
of $h_{\rm in} = 0.025 r_{\rm in}$.  Additionally, we have
assumed $H/r =0.03(r/r_{\rm in})^\beta$ with $\beta =0.3$ so that
$(H/r)_{\rm out} = 0.3$.   Also, 
 $U(r) =g (2GM/r)^{1/2}$ with $g=2$ and $M=M_\odot$, which is twice the local escape speed.}
	\end{figure}
%%%%%%%%%%%%%%%%%%%%%%%%%%%%%%%%%%%
 
       An estimate of the time $t_{\rm cr}$  at
 which $\hat{\Sigma}$  decreases to zero at $r_{\rm in}(t=0)$
 follows from Eqns. 13 and 14 assuming $ut \gg r_{\rm in}$,
 \begin{equation}
 t_{\rm cr} \approx {r_{\rm in}(0) \over u}  \left({ \dt{M}_{d0}~ g \sqrt{2} ~v_{Ki}
 \over \dt{M}_w~ (H/r)_i~ v_w}\right)^{1 \over \beta+1/2}~.
 \end{equation}
   Formally, $t_{\rm cr}$ is independent of $r_{\rm in}$.
 The dominant factors determining this time-scale are
 $u$ and $[v_w\dt{M}_w/(v_{Ki}\dt{M}_{d0})]$ with, for example,
 $t_{\rm cr} \propto  u^{-1}[v_{Ki}\dt{M}_{d0}/(v_w\dt{M}_w)])^{1.43}$ for  $\beta =0.3$.   
      The physical interpretation of Eqn. 15 is that the time-scale
 $t_{\rm cr}$ is larger than viscous accretion time-scale
 $r_{\rm in}(0)/u$ by a dimensionless factor which scales
 as a fractional power of
 the ratio of the momentum flux of the expelled disc matter
 $\dt{M}_{d} v_K$ to the momentum flux of the 
 wind $\dt{M}_w v_w$.

    Figure 3 shows the behavior of $\hat{\Sigma}(r,t)$
 for an illustrative case where 1) $t>0$, 2) the
 stellar wind has $\dt{M}_w = 10^{-10} M_\odot/$yr and $v_w=10^3$ km/s, 
 3) the initial disc accretion rate is $\dt{M}_{d0} =10^{-8}
 M_\odot$/yr, and 4) the viscosity coefficient is $\alpha =0.01$.
     A  central hole in the disc appears at $t/t_{\rm in}\approx 7.8\times10^3$
 when $\hat{\Sigma}(r_{\rm in},t)=0$  at $t\approx 2\times 10^6$ yr
 which is approximately equal to $t_{\rm cr}$ of Eqn. 15.

      After a hole starts to form for  $t> t_{\rm cr}$, we can
 rewrite the normalizations to  $r_{\rm in}$ in terms of
 normalizations to $r_{\rm out}=$ const.    For example,
 $H/r =(H/r)_{\rm out}(r/r_{\rm out})^\beta$.    In this
 way we find for $t>t_{\rm cr}$,
 \begin{equation}
 {r_{\rm hole}(t) \over r_{\rm in}(0)} \approx
1+ 0.56\left({t-t_{\rm cr} \over t_{\rm in}}\right)^{1.25}~,
 \end{equation}
 where $t_{\rm in} = r_{\rm in}(0)/u$.   Evidently, the
 hole expands rapidly for $t>t_{\rm cr}$.
 
 \section{Influence of coronal mass ejections}

   The foregoing has considered disc erosion by a steady stellar wind.
 Note however that significant   disc erosion can  arise
 the episodic component of the stellar wind, namely,  from frequent powerful
{\it coronal mass ejections} (CMEs).
Recent Kepler data provided new insights on the properties of stellar activity in magnetically active stars. 
   Specifically, they show the occurrence rates of flares with energies greater $10^{34}$ ergs referred to as {\it superflares}. 
   The data indicate that the occurrence rate of superflares from G-type stars follow the power-law relation with the flare's energy $E_f$ as
\begin{equation}
{dN \over dE_f} = k E_f^{-\alpha}~~{\rm events/s},
\end{equation}
where $\alpha \approx  2.1$ and $k \approx 2.7\times10^{32}$ in cgs units
(Aarnio, Matt, \& Stassun 2013). 
This characterization of stellar activity is important not only in terms of understanding the total radiative output from young stars, but also in terms of the mass output in the form of CMEs that usually accompany solar flares. CMEs cannot be directly observed from other solar-like stars except for possible type III and type IV bursts at decameter wavelength introduced by accelerated electrons as a CME propagates out from the solar/stellar corona (Boiko et al. 2012; Konovalenko et al. 2012; Massi et al. 2013). However, the frequency of CMEs from the young Sun and other active stars can be estimated from their association with solar/stellar flares. Recent SOHO/LASO and STEREO observations of energetic and fast ($\geq 500$ kms$^{-1}$) CMEs  from the
Sun show strong association with powerful solar flares (Yashiro \& Gopalswamy 2009; Aarnio et al. 2011). This empirical correlation provides a direct way to characterize CME frequencies of occurrence from statistics of solar and stellar flares via Eqn. (17). The total ejected mass in each solar CME  scales with the energy of the associated flare emitted in X-rays as
 \begin{equation}
M_{\rm CME} = (2.7 \pm 1.2)\times 10^{-3} E_f^{\beta}~~g~,
\end{equation}
where $\beta \approx 0.63 $ and $k_M \approx 2.7\times 10^{-3}$
in cgs units  (Aarnio et al. 2011). 
The solar CME masses vary between $10^{15}$ and  $5 \times 10^{16}$ g and contributes about $4\%$ of the total mass loss due to the solar wind. 

If we extrapolate this relation to stellar superflares on young stars with the maximum energy of $E=10^{36}$ erg, then the ejected mass per single event in a T Tauri star can be obtained by integrating Eqn. 18 over the occurrence rate from Eqn. 17  as
\begin{equation}
\dt{M}_{\rm CME} = \int_{E_{\rm min}}^{E_{\rm max}} dE_f M(E_f){dN \over dE_f}~,
\end{equation}
where $E_{\rm min,max}$ are discussed by Aarnio et al. (2013).
Given the uncertainty of the power-law index $\alpha$ in Eqn. 17, the calculated mass loss rate due to CMEs is  $\sim (3 -10)\times 10^{-10}  M_\odot$yr.$^{-1}$
(Drake et al. 2013; Aarnio et al. 2013).

   The disc erosion rate due to sporadic high velocity  CMEs with $\dt{M}_{\rm CME}$ comparable to the above considered steady wind $\dt{M}_w$ is expected to be
similar to the steady disc erosion rate.  A CME impacting the disc at a
distance $r$ with momentum $M_{\rm CME} v_{\rm CME}$ can eject disc matter $\Delta M_d$  with momentum $\Delta M_d~ v_{\rm esc}(r)$.

\section{Conclusions}%%%%%%%%%%%%%%%%%%%%%%%%
	This work develops an analytic model of protodiscs's gas and entrained dust erosion due to 
         high-velocity magnetized stellar winds.
    The presence of the magnetic field leads to Reynolds numbers sufficiently large to cause a strongly turbulent wind/disk boundary layer.  This boundary layer entrains and carries away the disc gas and entrained dust.
Strong  magnetized winds from young stars ($\lesssim 10^7$ yr) with discs are likely because the stars are known to rotate rapidly and  to be strongly magnetized.    
      The analytic model assumes a steady stellar wind with mass loss rate
 $\dt{M}_w \sim 10^{-10}M_\odot$ yr$^{-1}$ and velocity $v_w \sim
 10^3 $~km s$^{-1}$. 
       However,  in \S 4
 we discuss the  contribution to the disc erosion due to frequent
 powerful coronal mass ejections where the average mass
 loss rate in CMEs ($\dt{M}_{\rm CME}$)
 and velocities ($v_{\rm CME}$) have values comparable to those for the steady wind.

     Sample results for the evolution of the disc 
surface density $\hat{\Sigma}_d=\Sigma_d(r,t)/\Sigma(r,0)$ are
are shown in Figure 3.    The inner region of the disc surface
density decreases more rapidly than that at larger radii with
the result that a hole forms after a critical time $t_{\rm cr}$.
For the case shown, this time is about $2\times 10^6$ yr.
The critical time is proportional to the inverse accretion speed
in the disc $u=$ const times the ratio $\dt{M}_{d0}/\dt{M}_w$  raised to
a power larger than unity, where $\dt{M}_{d0}$ is the initial disc
accretion rate and $\dt{M}_w$ is the mass loss rate in the wind.
  The radius of the hole expands continuously with time.  
This is an important difference between wind erosion and photo-evaporation models where the inner hole is typically
less than $10$ au (Owen et al. 2011).    
     The possible role of
wind erosion for transition discs is complicated by the likely
presence of one or more planets at radii $ \lesssim 10$ au
(Espaillat et al. 2014).
   More realistic models would have $u$  dependent
on space and time, and $v_w$ and $\dt{M}_w$ dependent on time
as the star spins down.

   The  stellar wind disc erosion in the presence of  a radially distributed magneto-centrifugal disc wind (Blandford \& Payne 1982)   
remains to be investigated.    One possibility is that
any radially distributed poloidal magnetic field initially threading the disc
is advected inward to the disc/magnetosphere boundary where
it gives rise to a steady X-wind (Shu et al. 1994)
or to the episodic conical wind outflows found in {\it global}  MHD simulations (e.g.,  Romanova et al. 2009;  Lii et al. 2012, 2014; Dyda et al. 2013; Zanni and Ferreira 2013).  During the intervals when the conical
 wind is ``off'', the stellar wind can flow freely and impact the disc.
   Another possibility is that a radially distributed poloidal field $B_p(r)$ exists out to $r \gtrsim 10$ au
for the disc lifetime and gives rise to magnetically driven 
winds as found in the MHD  shearing-box simulations of Bai and Stone
(2013a, 2013b).  The mass loss in these winds could provide another
mechanism for disc dispersal (Armitage, Simon, \& Martin 2013).
   The stellar wind erosion discussed here is expected to dominate the magnetic disc winds if the ram pressure of the wind $\rho_w v_w^2$  is larger than  $B_p^2/8\pi$.

\section*{Acknowledgments}

   We thank Prof. J. P. Lloyd for  helpful discussions and an
anonymous referee for valuable criticism and suggestions  
on an earlier version of this work.
   This work  was supported in part by NASA grants NNX11AF33G,
NNX12AI85G, and NSF grant AST-1211318.


\begin{thebibliography}{1}
 
\bibitem[Aarnio et al.(2011)]{Aarnio2011}
Aarnio, A.N., Stassun, K.G., Hughes, W.J., \& McGregor, S.L. 2011,
Solar Phys., 268, 195

\bibitem[Aarnio et al. (2013)]{Aarnio2013}
Aarnio, Matt, S.P., \& Stassun, K.G. 2013,
Astron Nachr. 334, 77

\bibitem[Andrews \& Williams (2005)]{Andrews2005}
Andrews, S.~M.,  Williams, J.~P.  2005, ApJ, 631, 1134

\bibitem[Armitage \& Clarke (1996)]{Armitage1996}
Armitage, P.J., \& Clarke, C.J. 1996, MNRAS, 280, 458

\bibitem[Armitage (2010)]{Armitage2010a}
Armitage, P.~J.  2010, {\it The Astrophysics of Planet Formation}.
(Cambridge University Press: New York)

\bibitem[]{} Armitage, P.J., Simon, J.B., \& Martin, R.G. 2013,
ApJ, 778, L14

\bibitem[]{}
Bai, X.-N, \& Stone, J.M. 2013a, ApJ, 767, 30

\bibitem[]{}
Bai, X.-N, \& Stone, J.M. 2013b, ApJ, 769, 76

\bibitem[Belcher \& MacGregor (1976)]{Belcher1976}
Belcher J.~W., \& MacGregor K.~B. 1976, ApJ, 210, 498

\bibitem[Birnstiel et al. (2013)]{Birnstiel2013}
Birnstiel, T., Pinilla, P., Andrews, S.M., Benisty, M., \& Eercolano, B. 2013,  {\it Instabilities and Structures in Proto-Planetary Disks}, eds. Barge, P.,  and Jorda, L.,   EPJ Web of Conferences 46

\bibitem[Blandford \& Payne (1982)]{Blandford1982}
Blandford, R.D., \& Payne, D.G. 1982, MNRAS, 199, 883

\bibitem[Boiko et al. (2012)]{Boiko2012}	
	Boiko, A. I.,  Melnik, V. N.,  Konovalenko, A. A.,  Abranin, E. P., Dorovskyy, V. V., \& Rucker, H. O. 2012, Advances in Astron. and Space Phys., V.  2, 76

\bibitem[Bouvier et~al. (1993)]{Bouvier1993}
Bouvier, J.,  Cabrit, S.,  Fern\'{a}ndez, M.,  Mart\'{\i}n E.~L.,  \&  Matthews,  J.~M.  1993,  A\&A, 272, 176

\bibitem[Bouvier et al. (2014)]{Bouvier2014} Bouvier, J., Matt, S.P., Scholz, A., Stassun, K.G., \& 
Zanni, C. 2014, in Protostars \& Planets VI (University of
Arizona Press, eds H. Beuther, R. Klessen, K. Dullemond, \&
Th. Henning) (arXiv: 1309.7851v1)

\bibitem[Braginskii  (1965)]{Braginskii1965}
Braginskii, S.~I.  1965, {\it Review of Plasma Physics}, 
Vol.~1, (Consultants Bureau: New York)



\bibitem[Clarke et al. (2001)]{Clark2001}
Clarke, C. J., Gendrin, A., \&  Sotomayor, M.  2001,
MNRAS, 328, 485

\bibitem[Dyda et al. (2013)]{Dyda2013}
Dyda, S.,  Lovelace, R. V. E., Ustyugova, G. V., Lii, P. S.,  Romanova, M. M., \& Koldoba, A. V. 2013, MNRAS,  432, 127




  
 \bibitem[Espaillat et al. (2007)]{Espaillat2007}
 	Espaillat, C.,  Calvet, N., D'Alessio, P., Bergin, E.,  Hartmann, L., Watson, D., Furlan, E.,  Najita, J.,  Forrest, W.,  McClure, M.,  Sargent, B.,  Bohac, C., \& Harrold, S.T.  2007, ApJ, 664,  L111
	
\bibitem[Espaillat et al. (2014)]{Espaillat2014} Espaillat, C., 
Muzerolle, J., Najita, J., Andrews, S., Zhu, Z., Calvet, N., Kraus, S. Hashimoto, J., Kraus, A., \& D'Alessio, P.  2014, in Protostars \& Planets VI (University of Arizona Press, eds H. Beuther, R. Klessen, K. Dullemond, \& Th. Henning) (arXiv: 1402.7103v1)	
	
 



\bibitem[Hartmann et~al. (1998)]{Hartmann1998b}
Hartmann, L.,  Calvet, N.,  Gullbring, E., \&   D'Alessio, P. 1998, 
    ApJ, 495, 385

\bibitem[Hollenbach et al. (2000)] {Hollenbach2000}
Hollenbach, D.~J.,  Yorke, H.~W.,    \& Richstone, D.  2000, in
{\it Protostars and Planets IV}, eds. Mannings V.,  Boss
  A.~P., \&  Russell S.~S., 
(Univ. Arizona Press: Tuscon),  p.~406




\bibitem[Johns-Krull \&
  Valenti (2000)]{Johns-Krull2000}
Johns-Krull, C.~M., \&  Valenti, J.~A.  2000, in
{\it Stellar Clusters and Associations: Convection, Rotation,
  and Dynamos}, eds.
 Pallavicini R.,  Micela G., \&
  Sciortino S.,   Proceedings from ASP Conference, Vol. 198. Vol.~198,
  {Measurements of stellar magnetic fields}.
San Francisco, CA, pp 371--380


\bibitem[Johns-Krull  (2007)]{Johns-Krull2007}
Johns-Krull, C.~M.  2007,  ApJ, 664, 975

\bibitem[Konigl (1991)]{Konigl1991} K\"onigl, A. 1991, ApJ, 370, L39

\bibitem[Konovalenko et al. (2012)]{Konovalenko2012}
Konovalenko, A. A.,  Koliadin, V. L., Boiko, A. I., Zarka, Ph., Griessmeier, J.-M., Denis, L., Coffre, A., Rucker, H. O.,  Zaitsev, V. V., Litvinenko, G. V., Melnik, V. N., Stanislavsky, A. A., Stepkin, S. V., Mukha, D. V.,  Brazhenko, A.,  Leitzinger, M., Odret, P., \& Scherf, M.
 2012, European Planetary Science Congress (23-28 September, 2012, Madrid,
 Spain), vol. 7

\bibitem[Lii et al. (2012)]{Lii2012} Lii, P., Romanova, M.M.,
\& Lovelace, R.V.E. 2012, MNRAS, 4200, 2020

\bibitem[Lii et al. (2014)]{Lii2014} Lii, P., Romanova, M.M.,
Ustyugova, G.V., Koldoba, A.V., \& Lovelace, R.V.E. 2014,
MNRAS, 441, 86

\bibitem[Long et al. (2005)]{Long2005}
Long, M., Romanova, M.M., \& Lovelace, R.V.E. 2005,  ApJ, 634, 1214


\bibitem[Lovelace et al. (1999)]{Lovelace1999}
Lovelace, R.V.E., Romanova, M.M., \& Bisnovatyi-Kogan, G.S. 1999,
ApJ, 514, 368

\bibitem[Lovelace et al. (2008)]{Lovelace2008}
Lovelace, R. V.~E.,  Romanova, M.~M.,  \&  Barnard, A.~W.  2008, MNRAS, 389, 1233



\bibitem[Matt \& Pudritz (2008a)]{Matt2008a}
Matt, S., \&  Pudritz, R.~E. 2008a, The ApJ, 678, 1109

\bibitem[Matt \& Pudritz (2008b)]{Matt2008b}
Matt, S.,  \& Pudritz, R.~E.  2008b,  ApJ,  681, 391




\bibitem[Michel (1969)]{Michel1969}
Michel, F.~C.,  1969,  ApJ, 158, 727

\bibitem[Massi et al. (2013)]{Massi2013}
Massi, M., Ros, E., Boboltz, D., Menten, K. M., Neidhšfer, J., Torricelli-Ciamponi, G., \& Kerp, J. 2013, Memorie della Societa Astronomica Italiana, v.84, p.359 

\bibitem[Owen et al. (2012)]{Owen2012}
Owen, J. E.,  Clarke, C.  J., \& Ercolano, B.  2012,
MNRAS, 422, 1880

\bibitem[Pietu et al. (2006)]{Pietu2006}
Pi\'etu, V.,  Dutrey, A., Guilloteau, S., Chapillon, E., \& Pety, J.  2006,
A\&A, 430, L43

\bibitem[Regaly et al. (2012)]{Regaly2012}
Reg\'aly, Zs.,  Juh‡sz, A., S‡ndor, Zs., \& Dullemond, C. P.  2012,
MNRAS, 419, 1701

\bibitem[Richling \&
  Yorke (1997)]{Richling1997}
Richling, S., \& Yorke, H.~W.  1997,  A\&A, 324, 317

\bibitem[Romanova et~al. (2005)]{Romanova2005}
Romanova, M.~M.,  Ustyugova, G.~V.,  Koldoba, A.~V.,  \&  Lovelace, R. V.~E.  2005, ApJ, 635, L165

\bibitem[Romanova et al. (2009)]{Romanova2009} Romanova, M.M.,
Ustyugova, G.V., Koldoba, A.V., \& Lovelace, R.V.E. 2009, 
MNRAS, 399, 1802
  
  \bibitem[Roy \& Blottner (2006)]{Roy2006} Roy, C.J., \& Blottner, F.G. 2006, Prog. in Aerospace Sciences, 42, 469



\bibitem[Schlichting (1968)]{Schlicting1968} Schlichting, H. 1968, {\it Boundary-Layer Theory}, (McGraw-Hill: New York), ch. 23

\bibitem[Shakura \& Sunyaev (1973)]{Shakura1973}
Shakura, N. I., \& Sunyaev, R. A. 1973, A\&A, 24, 337

\bibitem[Shu et al. (1994)]{Shu1994}
Shu, F., Najita, J., Ostriker, E., Wilkin ,F., Ruden, S., \& Lizano S. 1994, ApJ, 429, 781

\bibitem[Skrutskie et~al. (1990)]{Skrutskie1990}
Skrutskie,  M.~F.,  Dutkevitch,  D.,  Strom, S.~E.,  Edwards, S.,   \&  Strom, K.~M. 1990,  AJ, 99, 1187



\bibitem[Ustyugova et~al. (2006)]{Ustyugova2006}
Ustyugova,  G.~V.,  Koldoba, A.~V.,  Romanova,  M.~M., \&   Lovelace, R. V.~E.,  2006, ApJ, 646, 304

\bibitem[Weber \& Davis (1967)]{Weber1967}
Weber, E.~J.,  \& Davis,  L.  1967,  ApJ, 148, 217

\bibitem[Wolk \&
  Walter (1996)]{Wolk1996}
Wolk, S.~J., \& Walter, F.~M.  1996,  AJ, 111, 2066

\bibitem[Yashiro \& Gopalswamy]{Yashiro2009}
Yashiro, S., \& Gopalswamy, N. 2009, in IAU Symp. 257, Universal Heliophysical Processes, ed. N. Gopalswamy \& D. F. Webb (Cambridge: Cambridge Univ. Press), 233

\bibitem[Zanni \& Ferreira (2013)]{Zanni2013}
Zanni, C., \& Ferreira, J. 2013, A\&A, 550, A99

\bibitem[Zsom et~al. (2010)]{Zsom2010}
Zsom,  A.,  Ormel, C.~W.,  G\"{u}ttler,  C.,  Blum, J., \&   Dullemond, C.~P.  2010,
  A\&A, 513, A57


\end{thebibliography}
\end{document}